\documentclass[11pt]{article}
\usepackage[utf8]{inputenc}

\usepackage{fullpage}
\usepackage{url}

\newcommand{\IGNORE}[1]{}

\usepackage{amsmath, amssymb, amsthm}

\newtheorem{theorem}{Theorem}[section]
\newtheorem{proposition}[theorem]{Proposition}
\newtheorem{lemma}[theorem]{Lemma}

\newtheorem{claim}[theorem]{Claim}

\usepackage{subcaption}

\usepackage{tikz}
\usetikzlibrary{arrows.meta}
\tikzset{>={Latex[width=2mm,length=2mm]}}

\usetikzlibrary{decorations.pathreplacing}

\usetikzlibrary{calc}
\usetikzlibrary{fit, decorations.markings}
\tikzset{
    tree/.style={thin, draw=black},
    link/.style={thin, dashed, draw=black},
    leaf/.style={circle, thin, fill=none, draw=black},
    subroot/.style={minimum size = 1cm, rectangle, thin, fill=none, draw=black},
}
\tikzstyle{vertex}=[circle, draw, inner sep=0pt, minimum size=10pt]
\newcommand{\vertex}{\node[vertex]}

\newcommand{\opt}{\textsl{opt}}
\newcommand{\Real} {\mathbb{R}}

\newcommand{\ptn}{\ensuremath{\mathcal{P}}}
\newcommand{\ptnx}[2]{\ensuremath{\mathcal{P}^{(#1)}_{#2}}}
\newcommand\ptnsup{\ensuremath{\widehat{\Pi}}}

  \newcommand{\cost}{\mathit{cost}}
  \newcommand\linkset{\ensuremath{\mathcal{L}}}
  \newcommand\nonleaf{\ensuremath{\Psi}}
  \newcommand\comps{\mathit{{Comp}}}

\newcommand\tld[1]{\tilde{#1}}

\newcommand{\wgt}{\textup{wgt}}
\newcommand\lpicked{\ensuremath{\ell^*}}

\newcommand{\Pec}{\ensuremath{P_\text{EC}}}
\newcommand{\Pnc}{\ensuremath{P_\text{NC}}}
\newcommand{\hPnc}{\ensuremath{\hat{P}_\text{NC}}}

\newcommand{\ncomp}{\ensuremath{\nu}}
\newcommand{\hindex}[1]{\ensuremath{\eta_{#1}}}
\newcommand{\inc}[1]{\mathit{inc}^{#1}}

\title{On a Partition LP Relaxation for Min-Cost 2-Node Connected Spanning Subgraphs}

\date{}

\author{
Logan Grout\thanks{({\tt lcg58@cornell.edu})
{Operations Research and Information Engineering, 206 Rhodes Hall, Cornell University, Ithaca, NY 14853-3801, USA}}
\and
Joseph Cheriyan\thanks{({\tt jcheriyan@uwaterloo.ca})
{Department of Combinatorics \& Optimization, University of Waterloo, Canada}}
\and
Bundit Laekhanukit\thanks{({\tt lbundit@gmail.com})
{Institute for Theoretical Computer Science, Shanghai University of Finance and Economics, China}}
}

\begin{document}

\maketitle

\begin{abstract}
Our motivation is to improve on the best approximation guarantee
known for the problem of finding a minimum-cost 2-node connected
spanning subgraph of a given undirected graph with nonnegative edge
costs. We present an LP (Linear Programming) relaxation based on partition constraints.

The special case where the input contains a spanning tree of zero
cost is called 2NC-TAP.  We present a greedy algorithm for 2NC-TAP,
and we analyze it via dual-fitting for our partition LP relaxation.
\end{abstract}

\bigskip
\bigskip


\section{ \label{s:intro}  Introduction}
{
Two of the key problems in the area of approximation algorithms for network design
are the min-cost 2ECSS (2-edge connected spanning subgraph) problem, and
the
the min-cost 2NCSS (2-node connected spanning subgraph) problem.
The latter problem is as follows:
Given an undirected graph $G=(V,E)$ and nonnegative costs on the edges,
denoted by either $\cost\in\Real_+^{E}$ or $c\in\Real_+^{E}$,
find a minimum-cost spanning subgraph that is 2-node connected.
Throughout, we use $n:=|V|$ to denote the number of nodes of $G$.
{Recall that a graph is 2-node connected if it has $\geq3$ nodes, it is connected, and
the deletion of any one node leaves a connected graph.}
This problem is NP-hard, see \cite{GJ:book79}.

Approximation algorithms for the min-cost 2NCSS problem
have been studied for several decades, see \cite{FJ:sicomp81,FJ:tcs82,MMP:mp90}.
The best approximation guarantee known is~2
(see \cite{KN:algor03,FJW06}, though there are earlier references).
Most of these algorithms are based on LP relaxations, and
the analysis of the approximation guarantee shows that
the integrality ratio of the standard LP relaxation, the so-called set-pairs~LP,
is $\approx 2$ (see \cite{FJW06}).
The set-pairs~LP (for the min-cost 2NCSS problem) can be viewed
as a generalization of the cut~LP relaxation of the min-cost 2ECSS problem.
Recall that the cut~LP relaxation of the min-cost 2ECSS problem is:
\[ (P_{cut}) \quad \min \left\{ \sum_{e\in{E}} \cost(e) x_e \;:\; x(\delta(S))\geq{2},~~
	\forall\emptyset\subsetneq{S}\subsetneq{V};\quad 0\leq{x}\leq{1} \right\}.
\]
(See Section~\ref{s:prelims} for definitions and notation.)
The set-pairs~LP is obtained from the LP $(P_{cut})$ by adding
the following family of constraints for each node $w\in{V}$:
$\sum (x_e\;:\; e=uv\in{E}, u\in{S}, v\in{(V-w)-S}) \geq1,\;
	\forall\emptyset\subsetneq{S}\subsetneq{V-w}$.
Informally speaking, the additional constraints require that the deletion of any node $w$
results in a subgraph that is ``fractionally~connected".  
An interesting special case of the problem is
the Tree Augmentation Problem for 2-node connectivity (2NC-TAP),
where the instance $(G,\;c)$ contains a spanning tree of zero cost;
thus, $(V,\; \{e\in E\;:\; \cost(e)=0\})$ contains a spanning tree, denoted by $T$.
The edges of $E(G)-T$ are called \textit{links}.

Recent research in the area has focused on the design and analysis
of approximation algorithms that beat the 
``natural threshold'' of~2 for the approximation guarantee.
The Weighted Tree Augmentation Problem (WTAP) is a special case of
the min-cost~2ECSS problem 
where the instance $(G,\;c)$ contains a spanning tree of zero cost.
Cohen and Nutov \cite{CN:tcs13} designed an approximation algorithm
for a special case of WTAP with a guarantee of $(1+\ln2)<1.7$, based
on earlier work by Zelikovsky on the Steiner Tree problem \cite{Zel96}.
Recently, Traub and Zenklusen \cite{TZ21a} presented an approximation
algorithm for WTAP with guarantee $1+(\ln{2})+\epsilon < 1.7$, via
a so-called relative greedy algorithm. Subsequently, they improved
on the approximation guarantee for WTAP via local search, see \cite{TZ21b}.

Nutov \cite{N:waoa20} recently presented a 1.91-approximation algorithm for
\textit{unweighted} 2NC-TAP, thus beating the threshold~of~2;
this result does not use any LP relaxation.

One obstacle for beating the approximation threshold~of~2 for
the min-cost 2NCSS problem and for (weighted) 2NC-TAP is that
the set-pairs~LP relaxation
has integrality ratio $\geq 2-\epsilon$.
There is a simple family of examples of (unweighted) 2NC-TAP such that the 
integrality ratio of the set-pairs~LP relaxation is $\ge 2-\epsilon$
(where $\epsilon$ is a small positive number),
see Proposition~\ref{prop:set-pairs}.

One of the main contributions of this paper is
a stronger LP relaxation for the min-cost 2NCSS problem,
that we call the partition~LP relaxation.
The partition~LP relaxation~($P$) of the min-cost 2NCSS problem is
obtained from the cut~LP relaxation of the min-cost 2ECSS problem
by adding the family of constraints 
\[
\sum_{e\in e_{G-w}(\ptn)} x_{e} \geq |\ptn|-1,\;
	\forall \ptn \in \ptnsup(\comps(G^0-w))
\]
for each node $w\in{V}$,
where $G^0$ denotes $(V,\{e\in E\,:\,\cost(e)=0\})$, and
$\ptn$ denotes a partition of $V-w$
such that (the node-set of) each connected~component of $G^0-w$
is contained in one of the sets of $\ptn$.

Moreover, based on our partition~LP relaxation,
we design a simple greedy algorithm for (weighted) 2NC-TAP
that achieves approximation guarantee $H(\lambda-1)$,
where $\lambda$ denotes the maximum
length over all the tree-paths $T(\ell)$ of the links $\ell$ of the
instance, and $H(k)$ denotes the $k$-th harmonic number.
For example, our algorithm achieves an approximation guarantee of
$H(3)=\frac{11}{6}$ for instances of 2NC-TAP such that
each link $\ell$ induces a tree-path $T(\ell)$ of length $\leq4$, regardless
of the diameter of the initial tree.
(See Figure~\ref{fig:greedyTightHighDiam} for an example of such an instance.)
Fredrickson and J{\'{a}}J{\'{a}} \cite{FJ:sicomp81} showed that
such instances of 2NC-TAP are NP-hard.
Our algorithm and analysis are based on two well-known results:
(1)~the greedy algorithm for SCP (the Set Covering Problem) and 
its analysis via ``dual fitting"
(i.e., charging the cost incurred by the greedy algorithm to
a feasible solution of the dual of the standard LP relaxation), and
(2)~the analysis of the greedy algorithm for the minimum spanning tree (MST) problem
via ``dual fitting" with respect to the well-known partition~LP formulation of MST,
see \cite{C:orl89}, \cite[Sec.~2]{KPT:mp11}.

Recall that the input to SCP consists of a set $U$ of $m$ points $p_1,\dots,p_m$,
$n$ subsets $S_1,\dots,S_n$ of $U$, and
a non-negative weight $\cost(S_j)$ for each subset $S_j,\;j=1,\dots,n$.
The goal is to find a minimum weight collection of the subsets $S_j$
that contains $U$.
The standard LP relaxation has a variable $x_j$ for each subset $S_j$,
and a linear constraint for each point $p_i$:
$\min \{ \sum \cost(S_j) x_j ~:~
\sum_{S_j\,:\,S_j\owns{p_i}} x_j\geq1,\;\forall{p_i}\in{U};~ x\geq0 \}$.
The dual LP has a non-negative variable $y_i$ for each point $p_i\in{U}$:
$\max \{ \sum_{p_i\in{U}} y_i ~:~ \sum_{p_i\in{S_j}} y_i\leq\cost(S_j),\;j=1,\dots,n;~ y\geq0 \}$.
The greedy algorithm iteratively picks a subset of the minimum cost-coverage ratio,
where the cost-coverage ratio of a set $S_j$ (at any time $t$ in the execution)
is $\cost(S_j)/|S_j^{(t)}|$, where $|S_j^{(t)}|$ denotes
the number of as-yet-uncovered points in $S_j$. It is well known that if each set
$S_j$ in an instance of SCP contains $\leq k$ points, then the greedy algorithm for 
SCP achieves an approximation guarantee of $H(k)$.

We note that the connection to SCP is obvious for the well-known
(weighted) Tree Augmentation Problem for 2-edge connectivity (WTAP).
The input for WTAP is the same as for 2NC-TAP, namely, $G, c, T$
(where $T$ is a spanning tree of $G$ of cost zero),
and the goal is to find a set of links $F$ of minimum cost
such that $T\cup{F}$ is 2-edge connected.
To view a TAP~instance as an SCP~instance, let $U = T$,
thus, the points of the SCP~instance correspond to the edges of $T$,
and let the subsets $S_j$ of the SCP~instance correspond
to the tree-paths $T(\ell)$ of the links $\ell$ of the TAP~instance.
For example, consider a TAP~instance on $K_4$ (the complete graph on four nodes)
where $T$ is a claw (the star $K_{1,3}$ with three leaves).
The SCP~instance has three points (that map to the three edges of the claw),
and it has three subsets $S_j$ of size two each
(corresponding to the three links of $E(K_4)-T$).
Clearly, the greedy algorithm for SCP gives an approximation algorithm for
WTAP with a guarantee of $H(\lambda)$.
Based on this mapping between WTAP and SCP, Cohen and Nutov \cite{CN:tcs13}
designed an approximation algorithm for a special case of WTAP with a guarantee of $(1+\ln2)<1.7$.
The major component of their algorithm is a local improvement algorithm
for SCP whose running time depends on the structure of the instance, and,
an initial feasible solution.
Their analysis of the running time of their algorithm relies on
some key properties of WTAP that do \textit{not} apply to 2NC-TAP.
Cohen and Nutov start by rooting the tree $T$ at an arbitrary node, and then,
in polynomial time, they find  a 2-approximate feasible solution to WTAP
that consists only of so-called up-links;
a link $\ell$ of a rooted tree is called
an up-link if it connects a node and its ancestor.
Such feasible solutions do not exist for 2NC-TAP.
(To illustrate this point, consider the above example on $K_4$, and
suppose that the root $r$ of the claw is the non-leaf node;
to ensure that $K_4-\{r\}$ is connected,
the solution has to pick two links that are not up-links.)
Secondly, for any feasible solution $F$ of WTAP,
Cohen and Nutov show (via the so-called shadow-complete assumption) that
there is an equivalent feasible solution $F'$ such that
the family of tree-paths $\{ T(\ell)\;:\; \ell\in{F'} \}$ is
pair-wise (edge) disjoint;
that is, every edge of the tree $T$ is contained in
exactly one of the tree-paths $T(\ell),\;\ell\in{F'}$.
Clearly, this property does not apply to 2NC-TAP.
(To illustrate this point, again consider the above example on $K_4$, and
let $r$ be the non-leaf node of the claw;
to ensure that $K_4-\{r\}$ is connected,
any solution $F$ has to pick at least two links, hence,
one of the edges of the claw $T$
is contained in two of the tree-paths $T(\ell),\;\ell\in{F}$.)

Some of the high-level ideas behind our analysis of the greedy algorithm for 2NC-TAP
are as follows.
We map an instance of 2NC-TAP to an instance of SCP by
mapping the relevant partitions $\ptn$
(such that our partition LP has a constraint for $\ptn$) to the points $p_i$ of SCP, and
mapping the links to the subsets $S_j$ of SCP.
If every link covers $\leq k$ relevant partitions,
then the approximation guarantee of $H(k)$ would follow immediately
(from the analysis of the greedy algorithm for SCP).
But, a link $\ell$ could be incident to many relevant partitions
(since the variable $x_{\ell}$ could occur in $2^{\theta(n)}$ partition constraints).
Informally speaking, we bypass this difficulty as follows:
we maintain a current partition $\ptn_u^i$ for each iteration $i$ and
each non-leaf node $u$ of $T$, and
we fix the ``scaled'' dual variable $y_{\ptn_u^i}$ of the partition $\ptn_u^i$ to be
the \textit{difference} between
the cost-coverage ratio of the link that covers $\ptn_u^i$
	(for the first time in the execution)
and the cost-coverage ratio of the link that covers the ``previous'' partition $\ptn_u^{(i-1)}$
	(for the first time in the execution).
To derive the approximation guarantee,
we need to show that the dual solution is feasible, and
it ``recovers'' the cost of the links picked by the greedy algorithm (up to a factor of $H(k)$).
This follows because
(1)~the cost-coverage ratios of the links are non-decreasing over the execution, 
(2)~for any ``picked link'' $\ell$, and any non-leaf node $u$ (of $T$) such that
	$\ell$ covers the current partition $\ptn_u^j$,
	the sum of the ``scaled'' dual variables of
	the sequence of partitions $\ptn_u^1,\dots,\ptn_u^j$
	``telescopes'' to the cost-coverage ratio of $\ell$, and
(3)~the ``scaled'' dual objective value $\sum_{\ptn} (|\ptn|-1)y_{\ptn}$
is equal to the sum of the costs of the links picked by the greedy algorithm.

For example, suppose that $T$ is a star; thus,
$\lambda= \max\{|T(\ell)|\;:\:\ell\text{ is a link}\}=2$.
Then, our greedy algorithm for 2NC-TAP is the same as
Kruskal's MST (minimum spanning tree) algorithm applied
to the subgraph induced on the leaves of $T$,
and the dual solution found by our algorithm is the same
as the dual solution found by the algorithms of \cite{C:orl89}, \cite[Sec.~2]{KPT:mp11}.

We mention that some of the results and constructions of this paper
have appeared in preliminary form
in the thesis of the first author, see \cite{grout:thesis20}.
}


\section{ \label{s:prelims}  Preliminaries}

This section has definitions and preliminary results.
Our notation and terms are consistent with \cite{Diestel:book10} or \cite{Frank:book11},
and readers are referred to those texts for further information.

For a positive integer $k$, we use $[k]$ to denote the set $\{1,\dots,k\}$.
We denote the $k$-th harmonic number by $H(k)$.

Let $G=(V,E)$ be a (loop-free, simple) graph with non-negative costs on the edges.
We take $G$ to be the input graph, and
we use $n$ to denote $|V(G)|$.
We denote the cost of an edge $e$ of $G$ by $\cost(e)$.
For a set of edges $F\subseteq E(G)$, $\cost(F):=\sum_{e\in F}\cost(e)$,
and for a subgraph $G'$ of $G$, $\cost(G'):=\sum_{e\in E(G')}\cost(e)$.

A multi-graph $H$ is called $k$-edge connected if $|V(H)|\ge2$ and for
every $F\subseteq E(H)$ of size $<k$, $H-F$ is connected.
A multi-graph $H$ is called $k$-node connected if $|V(H)|>k$ and for
every $S\subseteq V(H)$ of size $<k$, $H-S$ is connected.
We use the abbreviations \textit{2EC} for ``2-edge connected,'' and
\textit{2NC} for ``2-node connected.''

We use the standard notion of contraction of an edge, see \cite[p.25]{Schrijver:book03}:
Given a multi-graph $H$ and an edge $e=vw$,
the contraction of $e$ results in the multi-graph $H/(vw)$ obtained from $H$
by deleting $e$ and its parallel copies and identifying the nodes $v$ and $w$.
(Thus, every edge of $H$ except for $vw$ and its parallel copies
is present in $H/(vw)$; we disallow loops in $H/(vw)$.)

For a graph $H$ and a set of nodes $S\subseteq V(H)$,
$\delta_H(S)$ denotes the set of edges that have one end~node in
$S$ and one end~node in $V(H)-S$.
(We omit subscripts such as $H$, when there is no danger of ambiguity.)
Moreover, $H[S]$ denotes the subgraph of $H$ induced by $S$, and
$H-S$  denotes the subgraph of $H$ induced by $V(H)-S$.
For a graph $H$ and a set of edges $F\subseteq E(H)$,
$H-F$ denotes the graph $(V(H),~E(H)-F)$.
We may use relaxed notation for singleton sets, e.g.,
we may use $H-v$ instead of $H-\{v\}$.
We may not distinguish between a subgraph and its node~set;
for example, given a graph $H$ and a set $S$ of its nodes, we use
$E(S)$ to denote the edge~set of the subgraph of $H$ induced by $S$.

For a spanning tree $T$ and a link $\ell$, we use
$T(\ell)$ to denote the path of $T$ between the two end~nodes of $\ell$.

\subsection{The min-cost 2NCSS problem}

Given an undirected graph $G$
and nonnegative edge costs $c\in\Real_+^{E}$,
the algorithmic goal in the min-cost 2NCSS problem
is to find a 2NC spanning subgraph of minimum cost.
(For notational convenience, we may denote an instance by $G$ instead of $(G,\;c)$.)
This problem is NP-hard.
We assume that the input graph $G$ is 2NC.
For any instance $H$, we denote the minimum cost of a 2-NCSS of $H$ by $\opt(H)$.
When there is no danger of ambiguity, we use $\opt$ rather than $\opt(H)$.

\subsection{Partitions}

A partition $\ptn$ of a ground set $W$ is a family of sets of $W$
such that each element of $W$ belongs to exactly one set of $\ptn$.
The number of sets in $\ptn$ is denoted by $|\ptn|$.

A partition is called \textit{proper} if it consists of non-empty sets.
A partition is called \textit{trivial} if it consists of a single set, namely, $W$.
A partition that consists of singleton sets is called a \textit{point~partition} of $W$.
For example, if $W=[4]$, then $\ptn=\{ \{1\}, \{2\}, \{3\}, \{4\} \}$
is a point~partition of $W$.
Let $H=(W,F)$ be a graph, and let $\ptn$ be a partition of $W$.
An edge $e$ of $H$ is said to \textit{cross} $\ptn$ if the two end~nodes of
$e$ are in different sets of $\ptn$. 
We use $e_H(\ptn)$ to denote the set of edges of $H$ that cross $\ptn$.
For example, if $H=K_4$ and $\ptn$ is the point~partition of $V(H)$,
then $e_H(\ptn)=E(H)$;
if $H=K_6$ and $\ptn$ is a partition of $V(H)$ into two sets of size~3,
then $e_H(\ptn)$ consists of 9~edges.

Let $T$ be a tree, and let $u$ be a non-leaf node of $T$.
Let $\pi(\comps(T-u))$ denote the partition of $V(T)-u$ induced by
the connected components of $T-u$; thus, for each connected component
$C$ of $T-u$, there is a set $V(C)$ in $\pi(\comps(T-u))$.
Let $\ptnsup(\comps(T-u))$ denote the set of partitions $\ptn$
of $V(T)-u$ such that $\pi(\comps(T-u))$ is a refinement of $\ptn$;
thus, any partition that can be obtained by ``merging" some of the sets
of $\pi(\comps(T-u))$ is an element of the set $\ptnsup(\comps(T-u))$.
For example, suppose that $T$ is a tree with three leaves
$v_1,v_2,v_3$ and one non-leaf node $u$ (of degree three).
Then, $\pi(\comps(T-u))=\{\{v_1\},\{v_2\},\{v_3\}\}$,
and the set $\ptnsup(\comps(T-u))$ consists of the five partitions
$\pi(\comps(T-u))$,
$\{\{v_1\}, \{v_2,v_3\}\}$,
$\{\{v_2\}, \{v_1,v_3\}\}$,
$\{\{v_3\}, \{v_1,v_2\}\}$,
$\{\{v_1,v_2,v_3\}\}$.

\subsection{ \label{s:partition-LP} A partition LP relaxation for the min-cost 2-NCSS problem}

The partition~LP relaxation~($P$) of the min-cost 2NCSS problem has been presented
in Section~\ref{s:intro}, and we recall it here for convenience.
($P$)~is obtained from the cut~LP relaxation of the min-cost 2ECSS problem
by adding the family of constraints 
\[
\sum_{e\in e_{G-w}(\ptn)} x_{e} \geq |\ptn|-1,\;
	\forall \ptn \in \ptnsup(\comps(G^0-w))
\]
for each node $w\in{V}$,
where $G^0$ denotes $(V,\{e\in E\,:\,\cost(e)=0\})$, and
$\ptn$ denotes a partition of $V-w$
such that (the node-set of) each connected~component of $G^0-w$
is contained in one of the sets of $\ptn$.

\subsection{Polynomial-time computations}

There are well-known polynomial~time algorithms for implementing
all of the basic computations in this paper, see \cite{Schrijver:book03}.
We state this explicitly in all relevant results,
but we do not elaborate on this elsewhere.


\section{ \label{s:greedyalgorithm}  A greedy algorithm for min-cost 2NC-TAP}

We present a greedy algorithm for 2NC-TAP that achieves an approximation
guarantee of $H(\lambda-1)$, where $\lambda$ denotes the maximum
length over all the tree-paths $T(\ell)$ of the links $\ell$ of the
instance.

\subsection{A primal and dual LP relaxation for 2NC-TAP}
We start by presenting the partition~LP relaxation for 2NC-TAP.
This LP has a non-negative variable $x_{\ell}$ for each link $\ell$ of the instance $G$.
We denote the set of links of $G$ by $\linkset(G)$.
For each non-leaf node $u$ of the given spanning tree $T$,
we have a family of partition constraints for the graph $G$.
We denote the set of non-leaf nodes of $T$ by $\nonleaf(T)$.
The family of partition constraints for $u$ is similar to the family
of partition constraints for the partition~LP for MST, and is as follows:
\\
for each partition $\ptn \in \ptnsup(\comps(T-u))$,
there is a constraint $x(e_{G-u}(\ptn))\geq|\ptn|-1$
(i.e., the sum of the $x$-values of the links crossing $\ptn$
is required to be at least $|\ptn|-1$).

The partition LP for 2NC-TAP, $(P)$, and the dual of this LP, $(D)$, are stated below.

\smallskip
\noindent
\framebox{
\begin{minipage}{0.96\textwidth}
\[	
	\text{(P)}\left\{
	\begin{array}{llll}
	\min & \sum_{\ell\in \linkset(G)} \cost(\ell) x_{\ell} \\
	\text{s.t.} & \sum_{\ell\in \linkset(G) \cap e_{G-u}(\ptn)} x_{\ell} & \geq |\ptn|-1 &
		\forall u\in\nonleaf(T),\\ 
		& & & \forall \, \ptn \in \ptnsup(\comps(T-u))\\
	& x_{\ell} & \geq 0 & \forall \ell \in \linkset(G).
	\end{array}
	\right. 
\]
\end{minipage}
}
\medskip

\noindent
\framebox{
\begin{minipage}{0.96\textwidth}
\[	
	\text{(D)}\left\{
	\begin{array}{llll}
	\max & \sum_{u\in\nonleaf(T)} \: \sum_{\ptn\in\ptnsup(\comps(T-u))}
		(|\ptn|-1) y_{\ptn} \\
	\text{s.t} &
		\sum_{u\in\nonleaf(T)} \quad
		\sum_{\substack{\ptn\in\ptnsup(\comps(T-u))\\
				\text{s.t.}\,\ell\in{e_{G-u}(\ptn)}}} y_{\ptn} & \leq \cost(\ell)
			& \forall \, \ell\in\linkset(G) \\
	& y & \geq 0.
	\end{array}
	\right. 
\]
\end{minipage}
}

\medskip
\noindent
\textbf{Remark}:
Let $x\in \Real_+^{\linkset(G)}$ be a feasible solution of $(P)$.
Let $\tld{x}\in \Real_+^{E(G)}$ be the vector such that
$\tld{x}_{\ell}=x_{\ell}$ for each link $\ell\in\linkset(G)$, and
$\tld{x}_e=1$ for each edge $e \in T$.
Then, $\tld{x}$ satisfies the cut~constraints for 2-edge connectivity.
To see this, consider any nonempty, proper set of nodes $S \subsetneq{V}$.
Clearly, $|T \cap \delta(S)|\geq1$, since $T$ is a spanning tree.
If $|T \cap \delta(S)|\geq2$, then $\tld{x}(\delta(S))\geq|T\cap\delta(S)|\geq2$.
Now, suppose that $|T\cap\delta(S)|=1$.
Let $vu$ be the unique edge in $\delta_T(S)$.
At least one of $v$ or $u$ must be a non-leaf node.
We may assume that $u$ is a non-leaf node and $u\not\in S$.
Observe that $S$ is (the node-set of) a connected component of
$T-vu$, hence, $S$ is (the node-set of) a connected component of $T-u$.
Thus, the partition $\{S,\;{(V-S)-\{u\}}\}$ is in $\ptnsup(\comps(T-u))$, and, moreover,
$x$ satisfies the constraint
  $\sum_{\ell\in \linkset(G) \cap e_{G-u}(\{S,\;{(V-S)-\{u\}}\})} x_{\ell} \geq 1$.
Hence, we have
    \[
	\tld{x}(\delta(S))
	\geq \tld{x}_{uv} + x(e_{G-u}(\{S,\;{(V-S)-\{u\}}\}))
	\geq 2.
    \]

The following result shows that the constraints $x\leq{1}$
are redundant, whenever $(P)$ has an optimal solution.

{
\begin{proposition} \label{propo:extremepoints-of-P}
The extreme points of ($P$) are contained in $[0,1]^E$.
\end{proposition}
\begin{proof}
Suppose there exists an extreme point $x$ such that for some $\hat{\ell} \in 
\linkset(G)$, $x_{\hat{\ell}} = 1 + \epsilon$ for some $\epsilon > 0$.
Let $\chi_{\hat{\ell}}$ be the standard basis vector corresponding to $\hat{\ell}$,
let $x' = x - \epsilon \chi_{\hat{\ell}}$, 
and let $x'' = x + \epsilon \chi_{\hat{\ell}}$.
Clearly, $x''$ is feasible for $(P)$.
But $x'$ is not feasible for $(P)$
(otherwise, $x$ would not be an extreme point).
Thus, there exists $u \in V$ and 
$\ptn \in \ptnsup\left(\comps(T-u)\right)$ 
such that $\hat{\ell} \in e_{G-u}(\ptn)$ and
    \[
        \sum_{\ell \in \linkset(G) \cap e_G(\ptn)} x'_{\ell}
	\quad<\quad |\ptn|-1 \quad\leq\quad
        \sum_{\ell \in \linkset(G) \cap e_G(\ptn)} x_{\ell}.
    \]
Note that $|\ptn| \geq 3$.
(Otherwise, if $|\ptn|=2$, then
$\sum_{\ell \in \linkset(G) \cap e_G(\ptn)} x'_{\ell} \geq x'_{\hat{\ell}} \geq
	1 = |\ptn|-1$.)
Let $v$ and $w$ be the end~nodes of $\hat{\ell}$,
let $S_v$ and $S_w$ be the sets of $\ptn$ that contain $v$ and $w$, respectively,
and let $\ptn_{(vw)}$ be obtained from 
$\ptn$ by replacing $S_v$ and $S_w$ by the union $S_v\cup{S_w}$.
Clearly, $|\ptn_{(vw)}| = |\ptn|-1$. Finally, note that
\begin{align*}
\sum_{\ell \in \linkset(G) \cap e_G(\ptn_{(vw)})} x_{\ell}  & \leq
    \left(\sum_{\ell \in \linkset(G) \cap e_G(\ptn)} x_{\ell} \right) - x_{\hat{\ell}} \\
&=  \left(\sum_{\ell \in \linkset(G) \cap e_G(\ptn)} x'_{\ell} \right) + \epsilon -
	x_{\hat{\ell}} \\
&<  ( |\ptn|-1 ) - (x_{\hat{\ell}} - \epsilon) ~~=~~
	( |\ptn|-1 ) - 1 ~~=~~ |\ptn_{(vw)}|-1.
\end{align*}
Thus, $x$ violates the constraint of $\ptn_{(vw)}$, and
this contradicts the assumption that $x$ is feasible for $(P)$.
\end{proof}
}

\subsection{A greedy algorithm for 2NC-TAP}

The algorithm applies a number of iterations,
and constructs a set $F$ of chosen links; initially, $F$ is the empty set.
Each iteration picks one link according to a greedy rule and adds it to $F$.
The algorithm stops when $T\cup{F}$ induces a 2-NCSS of $G$.
Moreover, the algorithm assigns a non-negative number, denoted $\wgt$,
to each partition in $\bigcup \{ \ptnsup(\comps(T-u))~:~u\in\nonleaf(T) \}$;
initially, $\wgt(\ptn)=0$ for each of these partitions $\ptn$.

For each iteration $i=1,2,\dots,$
let $F^i$ denote the set of links picked by the previous iterations $1,2,\dots,i-1$;
thus, $|F^i|=i-1$.
At (the start of) each iteration $i=1,2,\dots,$
for each non-leaf node $u$ of the given spanning tree $T$,
the algorithm maintains the so-called current partition $\ptn^i_u\in \ptnsup(\comps(T-u))$;
this partition corresponds to the connected components of $(T\cup{F^i})-u$
(i.e., the sets of $\ptn^i_u$ correspond to the node-sets of the
connected components of $(T\cup{F^i})-u$).

Informally speaking, the working of the first iteration is the same as
the first iteration of the greedy algorithm for the following SCP instance:
there are $|\nonleaf(T)|$ points $p_1,p_2,\dots,p_j,\dots$ corresponding to the
partitions $\ptn^i_u, u\in\nonleaf(T)$,
and there are $|\linkset(G)|$ sets $S_1,S_2,\dots,S_k,\dots$ corresponding
to the links $\ell\in\linkset(G)$;
moreover, the point $p_j$ is in set $S_k$ iff
$\ell_k \in e_G(\ptn^i_{u_j})$ where
$\ptn^i_{u_j}$ denotes the partition corresponding to $p_j$ and
$\ell_k$ denotes the link corresponding to $S_k$.

Formally speaking, for each link $\ell\in\linkset(G)$,
let $\inc{i}(\ell)$ denote the set of partitions $\ptn^i_u$ crossed by $\ell$, that is,
$\inc{i}(\ell) = \{\ptn^i_u~:~u\in\nonleaf(T),~~\ell\in{e_G(\ptn^i_u)}\}$.
The iteration picks a link $\lpicked$ among the links $\ell$ with
$\inc{i}(\ell)\not=\emptyset$ such that
	$\frac{ \cost(\lpicked) }{ |\inc{i}(\lpicked)| }$
is minimum.
Moreover, the iteration assigns the weight 
	$\frac{ \cost(\lpicked) }{ |\inc{i}(\lpicked)| }$
to each of the partitions in $\inc{i}(\ell)$;
thus, $\wgt(\ptn^i_u) = 
	\frac{ \cost(\lpicked) }{ |\inc{i}(\lpicked)| },~\forall u\in\nonleaf(T)~:~\lpicked\in{e_G(\ptn^i_u)}$.
Also, the iteration applies the required updates, namely,
$F^{i+1}:=F^i\cup\{\lpicked\}$, and
for each node $u\in\nonleaf(T)$,
$\ptn^{i+1}_u$ is obtained from $\ptn^i_u$ by merging the two sets of $\ptn^i_u$
that each contain an end~node of $\lpicked$.

\subsection{Analysis of the greedy algorithm for 2NC-TAP}

Consider an arbitrary node $u\in\nonleaf(T)$.
Let $\ncomp(u)$ denote the number of connected~components of $T-u$;
clearly, $\ncomp(u) = |\pi(\comps(T-u))|$,
where $\pi(\comps(T-u))$ denotes the partition of $V(T)-u$ induced
by the connected components of $T-u$.
Observe that $|\ptn^1_u|=\ncomp(u)$
(since $\ptn^1_u=\pi(\comps(T-u))$),
and after the $i$-th iteration of the greedy algorithm,
either $|\ptn^{i+1}_u|=|\ptn^i_u|$ or 
       $|\ptn^{i+1}_u|=|\ptn^i_u|-1$;
moreover, if iteration~$i$ is the last iteration (that picks a link), then $|\ptn^{i+1}_u|=1$.
Let $\ptnx{1}{u}, \ptnx{2}{u}, \dots, \ptnx{\ncomp(u)-1}{u}$ denote
the sequence of partitions of $\ptnsup(\comps(T-u))$ that are assigned
a positive weight during the running of the greedy algorithm, ordered
according to the sequence in which the weights are assigned by the algorithm;
thus, $\ptnx{1}{u}$ is the first partition of $\ptnsup(\comps(T-u))$ (in
the running of the greedy algorithm) that is crossed by the link
picked in an iteration,
     $\ptnx{2}{u}$ is the second partition of $\ptnsup(\comps(T-u))$ (in
the running of the greedy algorithm) that is crossed by the link
picked in an iteration, etc.

The dual solution (of the LP) corresponding to the run of the greedy algorithm
is defined as follows. 
For each node $u\in\nonleaf(T)$,
\begin{align*}
   y(\ptnx{1}{u}) &= \wgt(\ptnx{1}{u}) \\
   y(\ptnx{j}{u}) &= \wgt(\ptnx{j}{u}) - \wgt(\ptnx{j-1}{u}), \quad
	(j\in \{2,3,\dots, \ncomp(u)-1\}) \\
   y(\ptn) &= 0 \textup{~for all other partitions $\ptn\in\ptnsup(\comps(T-u))$.}
\end{align*}

\begin{lemma} \label{lem:ynonneg}
For each node $u\in\nonleaf(T)$ and each partition $\ptn_u\in\ptnsup(\comps(T-u))$,
we have $y_{\ptn_u} \geq0$.
\end{lemma}
\begin{proof}
Essentially, this follows from two facts:
(1)~suppose that at (the start of) the $i$-th iteration,
the current partition $\ptn^i_u$ is crossed by a link $\ell$;
then $\ell$ crosses $\ptn^h_u$ for all $h<i$
(that is, if $\ell$ is a ``candidate link'' w.r.t.\ the current
partition of $u$ in iteration~$i$, then in all previous iterations
$h=1,\dots,i-1,$ $\ell$ is a ``candidate link'' w.r.t.\ the partition
$\ptn^h_u$ of that iteration);
(2)~the ratios
	$\frac{ \cost(\lpicked) }{ |\inc{i}(\lpicked)| }$
cannot decrease during the running of the greedy algorithm (that
is, the ratio for an iteration is $\geq$ the ratio for any previous
iteration).

In more detail, for any $j\in \{2,3,\dots, \ncomp(u)-1\}$, we claim that
	$\wgt(\ptnx{j}{u}) \geq \wgt(\ptnx{j-1}{u})$.
This can be seen as follows.
Suppose that the greedy algorithm
assigned the weight of the partition $\ptnx{j-1}{u}$ in the $i_{j-1}$-th iteration,
thus, $\ptnx{j-1}{u} = \ptn^{i_{j-1}}_u$;
moreover, let $\ell_{j-1}$ denote the link picked by that iteration.
Similarly, suppose that the greedy algorithm
assigned the weight of the partition $\ptnx{j}{u}$ in the $i_{j}$-th iteration, and
let $\ell_{j}$ denote the link picked by that iteration.

Then $\ell_j \in e_G(\ptn^{i_{j-1}}_u)$;
moreover, for each node
$w\in\nonleaf(T)$ such that $\ptn^{i_{j}}_w \in \inc{i_j}(\ell_j)$ note that
      $\ell_j \in e_G(\ptn^{i_{j-1}}_w)$
(that is, if $\ell_j$ crosses the partition $\ptn^{i_{j}}_w$ of
a non-leaf node $w$, then $\ell_j$ crosses the partition $\ptn^{i_{j-1}}_w$).
Hence, the ratio for the link $\ell_j$ in the $i_{j-1}$-th iteration,
	$\frac{ \cost(\ell_j) }{ |\inc{i_{j-1}}(\ell_j)| }$,
is $\leq$ the ratio for the link $\ell_j$ in the $i_{j}$-th iteration.
Since the greedy algorithm picked the link $\ell_{j-1}$ (rather than $\ell_j$) in the 
$i_{j-1}$-th iteration, we have
	$\frac{ \cost(\ell_{j-1}) }{ |\inc{i_{j-1}}(\ell_{j-1})| } \leq
	 \frac{ \cost(\ell_j) }{ |\inc{i_j}(\ell_j)| }$.
Hence, the ratio for the $i_{j-1}$-th iteration is $\leq$
       the ratio for the $i_{j}$-th iteration, and hence,
	$\wgt(\ptnx{j}{u}) \geq \wgt(\ptnx{j-1}{u})$.
\end{proof}

\begin{lemma} \label{lem:ydualfeasible}
$\frac{1}{H(\lambda-1)} y$ is a feasible solution to the dual~LP~$(D)$.
\end{lemma}

\begin{proof}
Consider an arbitrary link $\ell\in\linkset(G)$.
Recall that $T(\ell)$ denotes the path of $T$ between the two end~nodes of $\ell$,
and let $Q$ denote the set of internal~nodes of $T(\ell)$. Note that this implies $|Q| \leq \lambda$.
We claim that
	\[\sum_{u\in\nonleaf(T)} \quad
	\sum_{\{\ptn\in\ptnsup(\comps(T-u))\,:\,\ell\in{e_G(\ptn)}\}} y_{\ptn} \leq H(|Q|) \; \cost(\ell).\]

First, consider any node $u\in\nonleaf(T)-Q$;
thus, $u$ is not incident to $T(\ell)$,
For any partition $\ptn\in\ptnsup(\comps(T-u))$, note that $\ell$ does not cross ${\ptn}$,
hence, $\sum_{\{\ptn\in\ptnsup(\comps(T-u))\,:\,\ell\in{e_G(\ptn)}\}} y_{\ptn} = 0$.

Now, consider any node $u\in{Q}$, and consider the partitions of $\ptnsup(\comps(T-u))$
that have positive weights, namely, $\ptnx{1}{u}, \ptnx{2}{u}, \dots, \ptnx{\ncomp(u)-1}{u}$.
Observe that if $\ell$ crosses $\ptnx{j}{u}$, then $\ell$ also crosses each
of the partitions $\ptnx{1}{u},\dots,\ptnx{j-1}{u}$.
Let $\hindex{u}$ denote the highest index $j$ such that $\ell$ crosses $\ptnx{j}{u}$.
We have
\[\sum_{\{\ptn\in\ptnsup(\comps(T-u))\,:\,\ell\in{e_G(\ptn)}\}} y_{\ptn} =
	\sum_{j=1}^{\hindex{u}} y_{\ptnx{j}{u}} = \wgt(\ptnx{\hindex{u}}{u}).\]

Let $u_1,u_2,\dots,u_{|Q|}$ be an ordering of the nodes in $Q$
according to the reverse of the order in which
the greedy algorithm assigns weights to the partitions $\{\ptnx{\hindex{u}}{u}~:~u\in{Q}\}$;
that is, $\ptnx{\hindex{u_1}}{u_1}$ is the last of these partitions to be assigned a weight,
         $\ptnx{\hindex{u_2}}{u_2}$ is the second~last of these partitions to be assigned a weight, etc.
We have $\wgt(\ptnx{\hindex{u_j}}{u_j}) \leq \cost(\ell)/j$, for each $j=1,\dots,|Q|$,
because at the iteration when the greedy algorithms assigns the weight of $\ptnx{\hindex{u_j}}{u_j}$,
the partitions $\ptnx{\hindex{u_1}}{u_1}, \dots, \ptnx{\hindex{u_{j-1}}}{u_{j-1}}, \ptnx{\hindex{u_j}}{u_j}$
are crossed by $\ell$, hence the weight assigned in that iteration
  cannot exceed $\cost(\ell)/j$.
Hence, $\sum_{u\in{Q}} \wgt( \ptnx{\hindex{u}}{u} ) \leq
	(1+\frac12+\dots+\frac{1}{|Q|}) \; \cost(\ell) \leq H(|Q|) \; \cost(\ell)$.
Therefore,
	$$\sum_{u\in\nonleaf(T)} \quad
	\sum_{\{\ptn\in\ptnsup(\comps(T-u))\,:\,\ell\in{e_G(\ptn)}\}} y_{\ptn} =
	\sum_{u\in{Q}} \wgt( \ptnx{\hindex{u}}{u} ) \leq H(|Q|) \; \cost(\ell).$$
\end{proof}

\begin{theorem} \label{thm:greedyalg}
The cost of the set of links $\hat{F}$ returned by the greedy algorithm, $\cost(\hat{F})$, is
$\leq H(\lambda-1) \; \opt(P)$, where $\opt(P)$ denotes the optimal value of the LP~$(P)$.
\end{theorem}

\begin{proof}
The description of the greedy algorithm
and the definition of the weights of the partitions imply that
for each iteration~$i$ and the link $\ell$ picked in that iteration,
\[
\cost(\ell) = \sum \{ \wgt(\ptn^i_u) \;:\;
	u \hbox{~is an internal node of~} T(\ell) \hbox{~and~} \ptn^i_u\in\inc{i}(\ell) \};
\]
Furthermore, if $\wgt(\ptn^i_u)>0$ then $\ptn^{i+1}_u \neq \ptn^i_u$ because,
in the $i$-th iteration, the end~nodes of $\ell$ are in different sets of $\ptn^i_u$,
whereas, in the $(i+1)$-th iteration,
the end~nodes of $\ell$ are in the same set of $\ptn^{i+1}_u$.
Hence, we have
$\cost(\hat{F}) = \sum_{i=1}^{|\hat{F}|} \sum_{u\in\nonleaf(T)} \wgt(\ptn^i_u)$.

By the previous lemma, $\frac{1}{H(\lambda-1)}\; y$ is a feasible solution of the dual~LP~$(D)$,
hence, the objective value of this feasible solution is $\leq \opt(P)$.
We rewrite this objective value:
\[
\begin{array}{ll}
	& \frac{1}{H(\lambda-1)}\; \sum_{u\in\nonleaf(T)} \:
		\sum_{\ptn\in\ptnsup(\comps(T-u))} (|\ptn|-1) y_{\ptn}\\
	 &= \frac{1}{H(\lambda-1)}\; \sum_{u\in\nonleaf(T)} \:
		\Big( \sum_{j=1}^{\ncomp(u)-1} \wgt({\ptnx{j}{u}}) \Big) \\
	 &= \frac{1}{H(\lambda-1)}\; \sum_{u\in\nonleaf(T)} \: \sum_{i=1}^{|\hat{F}|}         \wgt(\ptn^i_u) \\
	 &= \frac{1}{H(\lambda-1)}\; \cost(\hat{F}).
\end{array}
\]
To derive the first two equations, consider any non-leaf node $u$, and note that
\[
\begin{array}{ll}
	\sum_{\ptn\in\ptnsup(\comps(T-u))} (|\ptn|-1) y_{\ptn} 
	&= \sum_{j=1}^{\ncomp(u)-1} (\ncomp(u)-j) y_{\ptnx{j}{u}} \\
	&= \sum_{j=1}^{\ncomp(u)-1} (\ncomp(u)-j) \wgt({\ptnx{j}{u}}) \\
	& \qquad -\sum_{j=1}^{\ncomp(u)-2} (\ncomp(u)-j-1) \wgt({\ptnx{j}{u}})\\
	&= \sum_{j=1}^{\ncomp(u)-1} \wgt({\ptnx{j}{u}})\\
	&= \sum_{i=1}^{|\hat{F}|} \wgt(\ptn^i_u).
\end{array}
\]

Therefore, $\cost(\hat{F}) \leq H(\lambda-1)\; \opt(P)$.
\end{proof}

{
\subsection{A tight example for the greedy algorithm for 2NC-TAP}

The example of Figure~\ref{fig:greedyTight} shows that our analysis
of the greedy algorithm (in Theorem~\ref{thm:greedyalg}) is tight when $\lambda=4$.
This can be generalized to any positive integer $\lambda\geq2$ as follows.
The problem instance has $T$ being a path with vertex set
$[\lambda+1]$ and edge set $\{\{k,k+1\} : k \in [\lambda]\}$.
For $k \in [\lambda-1]$ let $\ell_k = \{k,k+2\}$ and $\tld{\ell} = \{1,\lambda+1\}$. 
The link set is $\{\ell_k : k \in [\lambda-1]\} \cup \{\tld{\ell}\}$.
Finally, let the cost of the links be given by
$\cost(\ell_k) = \frac{1}{k}$ for $k \in [\lambda-1]$ and
$\cost(\tld{\ell}) = 1+\epsilon$.
For each non-leaf node $u$, note that $T-u$ has two connected components;
hence, the LP has a unique constraint of the form
$\sum_{\ell\in \linkset(G) \cap e_G(\ptn_u)} x_{\ell} \geq |\ptn_u|-1$,
where $\ptn_u$ is a partition of $V-u$ with $|\ptn_u|=2$.
For each $k \in [\lambda-1]$, note that $\ell_k$ crosses $\ptn_{k+1}$.
Moreover, the link $\tld{\ell}$ crosses each of these $\lambda-1$ partitions.
At the start of the $i$-th iteration,
the ratio $\frac{\cost(\tld{\ell})}{|\inc{i}(\tld{\ell})|} = \frac{1+\epsilon}{\lambda-1-(i-1)}$.
However, even after picking the first $i-1$ links,
there is still a link of cost 
$\frac{1}{\lambda-1-(i-1)} < \frac{1+\epsilon}{\lambda-1-(i-1)}$, namely,
$\ell_{\lambda-1-(i-1)}$,
so the greedy algorithm will pick that link.
Thus, the greedy algorithm finds a solution of cost
	$\sum_{i=1}^{\lambda-1}\frac{1}{\lambda-1-(i-1)} =
	\sum_{i=1}^{\lambda-1}\frac{1}{i} = H(\lambda-1)$.
Observe that an optimal solution has cost $1+\epsilon$ and consists of the link $\tld{\ell}$.

{
\begin{figure}[htb]
    \centering
    \begin{tikzpicture}[
        level 1/.style={sibling distance=30mm},level 2/.style={sibling distance=15mm}
    ]
        \node[thin,circle,draw=black] (root) {$v_3$}
            child { node[thin,circle,draw=black] (LR) {$v_2$} edge from parent[tree]
                child { node[thin,circle,draw=black] (L) {$v_1$} edge from parent[tree]
                }
            }
            child { node[thin,circle,draw=black] (RR) {$v_4$} edge from parent[tree]
                child { node[thin,circle,draw=black] (R) {$v_5$} edge from parent[tree]
                }
            };
        
        \draw [link] (L) .. controls ($(LR)+(-0.75,0.5)$) .. (root) node[midway,above] {$6$};
        \draw [link] (LR) -- (RR) node[midway,above] {$3$};
        \draw [link] (R) .. controls ($(RR)+(0.75,0.5)$) .. (root) node[midway,above] {$2$};
        \draw [link] (L) -- (R) node[midway,above] {$6+\epsilon$} node[midway,below] {$l^*$};
    \end{tikzpicture}
    \caption{ \label{fig:greedyTight} An instance of 2NC-TAP such that the greedy algorithm
	returns a solution of cost $\frac{11}{6}$ times the optimal cost.
	Edges indicated by solid lines have cost~0 and
	edges indicated by dashed lines are labelled with their costs.
	}
\end{figure}
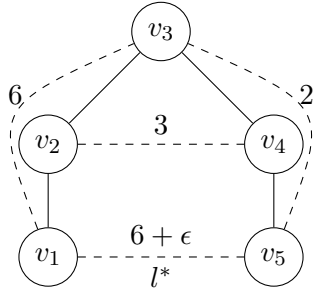
}

{
Using links of cost~0, one can easily string together multiple copies
of this example to obtain a graph of arbitrary diameter, as shown in
Figure~\ref{fig:greedyTightHighDiam}.  In particular, for $\lambda\geq3$,
if we have $k$ copies of the above example, and the $i$-th copy has
vertex set $v^{(i)}_{1}, \ldots, v^{(i)}_{\lambda+1}$, then adding
tree edges $v^{(i)}_1v^{(i+1)}_1$ and links $v^{(i)}_2v^{(i+1)}_2$,
of cost 0, for $i \in [k-1]$, results in an instance that has diameter
$\geq k+1$; note that $\lambda$ is still the maximum of the lengths
of the tree paths $T(\ell)$ defined by the links $\ell$.

{
\begin{figure}[htb]
    \centering
    \begin{tikzpicture}
        \vertex (11) at (0,0) {$v^{(1)}_1$};
        \vertex (12) at ($(11) + (1,0)$) {$v^{(1)}_2$};
        \vertex (13) at ($(11) + (2,0)$) {$v^{(1)}_3$};
        \vertex (14) at ($(11) + (3,0)$) {$v^{(1)}_4$};
        \vertex (15) at ($(11) + (4,0)$) {$v^{(1)}_5$};
       
        \draw[tree] (11) -- (12) -- (13) -- (14) -- (15);
       
        \draw [link] (11) .. controls ($(12)+(0,1)$) .. (13) node[midway,above] {$6$};
        \draw [link] (12) .. controls ($(13)+(0,1)$) .. (14) node[midway,above] {$3$};
        \draw [link] (15) .. controls ($(14)+(0,1)$) .. (13) node[midway,above] {$2$};
        \draw [link] (11) .. controls ($(13)+(0,-1)$).. (15) node[midway,above] {$6+\epsilon$};
        \vertex (21) at ($(11) + (0,3)$) {$v^{(2)}_1$};
        \vertex (22) at ($(21) + (1,0)$) {$v^{(2)}_2$};
        \vertex (23) at ($(21) + (2,0)$) {$v^{(2)}_3$};
        \vertex (24) at ($(21) + (3,0)$) {$v^{(2)}_4$};
        \vertex (25) at ($(21) + (4,0)$) {$v^{(2)}_5$};
       
        \draw[tree] (21) -- (22) -- (23) -- (24) -- (25);
       
        \draw [link] (21) .. controls ($(22)+(0,1)$) .. (23) node[midway,above] {$6$};
        \draw [link] (22) .. controls ($(23)+(0,1)$) .. (24) node[midway,above] {$3$};
        \draw [link] (25) .. controls ($(24)+(0,1)$) .. (23) node[midway,above] {$2$};
        \draw [link] (21) .. controls ($(23)+(0,-1)$).. (25) node[midway,above] {$6+\epsilon$};
        \vertex (31) at ($(21) + (0,3)$) {$v^{(3)}_1$};
        \vertex (32) at ($(31) + (1,0)$) {$v^{(3)}_2$};
        \vertex (33) at ($(31) + (2,0)$) {$v^{(3)}_3$};
        \vertex (34) at ($(31) + (3,0)$) {$v^{(3)}_4$};
        \vertex (35) at ($(31) + (4,0)$) {$v^{(3)}_5$};
       
        \draw[tree] (31) -- (32) -- (33) -- (34) -- (35);
       
        \draw [link] (31) .. controls ($(32)+(0,1)$) .. (33) node[midway,above] {$6$};
        \draw [link] (32) .. controls ($(33)+(0,1)$) .. (34) node[midway,above] {$3$};
        \draw [link] (35) .. controls ($(34)+(0,1)$) .. (33) node[midway,above] {$2$};
        \draw [link] (31) .. controls ($(33)+(0,-1)$).. (35) node[midway,above] {$6+\epsilon$};
        \draw[tree] (11) -- (21) -- (31);
        \draw[link] (12) .. controls ($(11)+(0.5,1.5)$) .. (22) node[midway,left]{$0$};
        \draw[link] (22) .. controls ($(21)+(0.5,1.5)$) .. (32) node[midway,left]{$0$};
    \end{tikzpicture}
    \caption{ \label{fig:greedyTightHighDiam}
	An instance of 2NC-TAP with diameter $\geq{k+1}$ (shown with $\lambda=4$ and $k=3$).
	Edges indicated by solid lines have cost~0 and
	edges indicated by dashed lines are labelled with their costs.
	An optimal solution uses the links of cost~0 and the links of cost $6+\epsilon$.
	The greedy algorithm returns a solution that uses all of the links except
	those of cost $6+\epsilon$.
	}
\end{figure}
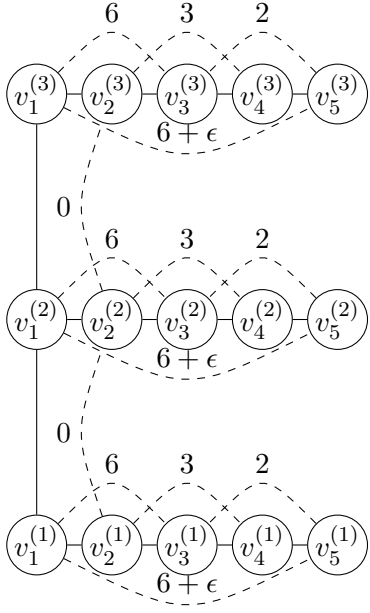
}
}
}


{
\section{ \label{s:intratio}  The integrality ratio of the partition~LP relaxation of 2NCSS}

In this section, we focus on the integrality ratio of the partition~LP relaxation of 2NCSS.
We show that the integrality ratio is $\leq2$;
this holds because the well-know set-pairs~LP for 2NCSS has integrality ratio $\approx{2}$,
and the set-pairs~LP is a relaxation of the partition~LP.
Next, we show (via a simple construction) that the integrality ratio
of the partition~LP relaxation of 2NCSS is $\geq$
the integrality ratio of the well-known cut~LP relaxation of 2ECSS.
The latter~LP is known to have integrality ratio $\geq \frac{3}{2}$ \cite{CKKK:orl08}
(in fact, the ratio $\frac{3}{2}$ is achieved on a family of instances of TAP,
the tree augmentation problem).

The instance of unweighted 2NC-TAP in Figure~\ref{fig:intGap} shows that
the partition~LP for 2NC-TAP has integrality ratio $\geq \frac{4}{3}$;
an optimal solution of the instance has cost~4,
whereas the partition~LP has a (fractional) solution of cost $3$.
This example has $\lambda=4$.
Theorem~\ref{thm:greedyalg} gives an upper-bound of $H(\lambda-1)=H(3)=\frac{11}{6}$
on the integrality ratio of any instance of 2NC-TAP with $\lambda=4$.
Possibly, the analysis of Theorem~\ref{thm:greedyalg}
could be improved for some special cases;
it is not clear whether an approximation ratio of $\frac{4}{3}$ can be proved
for instances of unweighted 2NC-TAP with $\lambda=4$, see \cite{DF97}.
(Duh and F\"{u}rer \cite{DF97} presented a $\frac{4}{3}$-approximation algorithm
for unweighted 3-SCP via semi-local optimization;
3-SCP is the special case of the Set Covering Problem where
$|S_j|\leq3$ for each of the sets $S_j$ of the instance.)

\begin{proposition} \label{prop:set-pairs}
The set-pairs~LP~relaxation for the min-cost 2NCSS problem has integrality ratio $\approx2$.
\end{proposition}
\begin{proof}
The upper-bound of~2 on the integrality ratio (for the set-pairs~LP)
follows from the analysis of the 2-approximation guarantee for the min-cost 2NCSS problem
relative to the set-pairs~LP by Fleischer et~al., see \cite[Theorems~3.13,~3.14]{FJW06}.
(In fact, Fleischer et~al.\ prove the 2-approximation guarantee for a more
general problem, namely, VC-SNDP with requirements of $\{0,1,2\}$
openly-disjoint paths between pairs of nodes;
the min-cost 2NCSS problem is a special case of VC-SNDP.)

A lower-bound of $2-\frac{\Theta(1)}{n}$ is implied by the following
well-known example:
Consider an instance of unweighted 2NC-TAP that
consists of the spanning tree $T = K_{1,n-1}$ (thus, $T$ is a star),
and $n-1$ (unit-cost) links that form a cycle on the leaves of $T$.
Any integer solution picks $n-2$ links, and has cost $n-2$.
There is a (fractional) solution $\hat{x}$ to the set-pairs~LP of cost $(n-1)/2$
that fixes $\hat{x}_{\ell}=\frac{1}{2}$ for each link $\ell$.
\end{proof}

\subsection{A transformation from 2ECSS to 2NCSS}

In this subsection,
we show that the integrality ratio of our partition LP relaxation
is $\geq{1.5}$ by giving a transformation from
TAP (the Tree Augmentation Problem for 2-edge connectivity) to
the min-cost 2NCSS problem
that preserves the integrality ratio.
There is a well-known construction for TAP that has integrality ratio $1.5$, see \cite{CKKK:orl08}.
In this subsection, we denote the cost of an edge $e$ by $c_e$ or $c'_e$.

\medskip

Let $G=(V,E)$ be a graph, and let each edge $e$ have a cost $c_e\in\Real$.
Let $\Pec(G)$ denote the feasible region of
the cut~LP relaxation of the min-cost 2ECSS problem:
\[ \min \left\{ \sum_{e\in{E}} c_e x_e \;:\; x(\delta(S))\geq{2},~~
	\forall\emptyset\subsetneq{S}\subsetneq{V};\quad 0\leq{x}\leq{1} \right\}.
\]

Let $\Pnc(G)$ denote the feasible region of
the partition~LP relaxation~($P$) of the min-cost 2NCSS problem,
see Section~\ref{s:partition-LP}.

The following well-known construction (inflation) maps
an instance $(G,\;c)$ of the min-cost 2ECSS problem
to an instance $(G',\;c')$ of the min-cost 2NCSS problem.
Each node $u$ of $G$ maps to a
distinct clique $C'_u$ on $\deg_G(u)$ nodes of $G'$
(that is, $C'_u$ is a complete graph on $\deg_G(u)$ nodes
and $C'_u,C'_w$ are node-disjoint for any two nodes $u,w\in{V(G)},u\not={w}$),
and each edge $vw$ of $G$ maps to an
edge $v'w'$ of $G'$ that has one end~node $v'$ in $C'_v$ and
has the other end~node $w'$ in $C'_w$
such that each node of a clique $C'_u$ of $G'$ is
incident to exactly one inter-clique edge;
moreover, $c'_{v'w'} = c_{vw},\:\forall{vw\in{E}}$,
and for each edge $e'$ of a clique $C'_u$ of $G'$, the cost $c'_{e'}$ is zero.
Let $F'=\bigcup_{u\in{V(G)}}E(C'_u)$;
thus, $F'$ consists of the edges of $G'$ that have
both end~nodes in the same clique $C'_u$ of $G'$, $u\in V(G)$.
Figure~\ref{fig:transform} illustrates this construction on an instance of TAP
(note that the Tree Augmentation Problem is
a special case of the min-cost 2ECSS problem).

\begin{proposition} \label{prop:intratio}
Let $\big(G=(V,E),~ c\in\Real^{E}\big)$ be an instance of the min-cost 2ECSS problem, and
let $\big(G'=(V',E'),~ c'\in\Real^{E'}\big)$ denote the instance of the min-cost 2NCSS problem
that is obtained from $(G,\;c)$ by the above construction.
The integrality ratio of the cut~LP for $(G,\;c)$
is the same as the
integrality ratio of the partition~LP for $(G',\;c')$.
\end{proposition}
\begin{proof}
Our proof is based on two claims.

\begin{claim} \label{claim:EcToNc}
For $x \in \Pec(G)$, define $x' \in \Real^{E'}$ as follows:
\[x'_{e'} = \begin{cases} x_{vw} & \text{ if $e' \in E'-F'$ and } e' = v'w'\\
                            1 & \text{ if } e' \in F' \end{cases} \]
Then $x' \in \Pnc(G')$ and $c'^\top x' = c^\top x$.
\end{claim}

\begin{claim} \label{claim:NcToEc}
Let $x' \in \Pnc(G')$ and define $x \in \Real^{E}$ as 
$x_{vw} =  x'_{v'w'} \quad \forall vw \in E$. 
Then $x \in \Pec(G)$ and $c^\top x = c'^\top x'$.
\end{claim}

Let $\hPnc(G') = \{x \in \Pnc(G') : x_e = 1,~ \forall e \in F'\}$.
Note that, if the partition~LP for $G'$ has an optimal solution,
then there exists an optimal solution in $\hPnc(G')$;
this holds because $c'_e = 0, \forall e\in{F'}$.
Claims~\ref{claim:EcToNc} and~\ref{claim:NcToEc} give us
a bijection $\varphi: \Pec(G) \to \hPnc(G')$ such that
$c^\top x = c'^\top \varphi(x)$ for all $x \in \Pec(G)$.
Furthermore, $\varphi$ maps integral vectors to integral vectors.

Let $x_*$ be an optimal (fractional) solution of the cut~LP relaxation
of the min-cost 2ECSS instance $(G,\;c)$, and
let $z_*$ be an optimal integral solution of the same LP
(thus, $z_*$ is a min-cost 2ECSS of $(G,\;c)$).
Similarly, let $x'_*$ be an optimal (fractional) solution of the partition~LP relaxation
of the min-cost 2NCSS instance $(G',\;c')$, and
let $z'_*$ be an optimal integral solution of the same LP
(thus, $z'_*$ is a min-cost 2NCSS of $(G',\;c')$).
Thus, we have
    $c^\top x_* = c'^\top x'_*$, because
	$\displaystyle
    \big( c^\top x_* = c'^\top \varphi(x_*) \geq c'^\top x'_*\big)  \text{~and~}
    \big( c^\top x_* \leq c^\top \varphi^{-1}(x'_*) = c'^\top x'_* \big)$.
Similarly, we have $c^\top z_* = c'^\top z'_*$.
Therefore, 
	$\displaystyle \frac{c'^\top z'_*}{c'^\top x'_*} = \frac{c^\top z_*}{c^\top x_*}.$
Hence, the two instances $(G,\;c)$ and $(G',\;c')$ have the same integrality ratios
with respect to their LP relaxations (namely, the cut~LP and the partition~LP).
\end{proof}

Figure~\ref{fig:transform} illustrates Proposition~\ref{prop:intratio} and our construction.
Figure~\ref{fig:tap}~(a) shows a TAP~instance from the family of TAP~instances
with integrality ratios converging to $\frac{3}{2}$, see \cite{CKKK:orl08}.
The application of our construction to this TAP~instance
results in the instance of the min-cost 2NCSS problem in Figure~\ref{fig:transform}~(b).
Moreover, the integrality ratio of this particular TAP~instance for the cut~LP relaxation
is the same as the
integrality ratio of the instance of the min-cost 2NCSS problem for the partition~LP relaxation.
}
{
\begin{figure}[htb]
    \centering
{
\begin{tikzpicture}[scale=0.75]
    \vertex (v_1) at (0,0) {};
    \vertex (v_2) at (0,1) {};
    \vertex (v_3) at (0,2) {};
    \vertex (w_11) at (-3,0) {${p_1}$};
    \vertex (w_12) at (3,0) {${q_1}$};
    \vertex (w_21) at (-3,1) {${p_2}$};
    \vertex (w_22) at (3,1) {${q_2}$};
    \vertex (w_31) at (-3,2) {${p_3}$};
    \vertex (w_32) at (3,2) {${q_3}$};
    \draw[] (w_11) -- (v_1) -- (w_12);
    \draw[] (w_21) -- (v_2) -- (w_22);
    \draw[] (w_31) -- (v_3) -- (w_32);
    \draw[dashed] (w_11) -- (w_21) -- (w_31) to[out=180,in=180] (w_11);
    \draw[dashed] (w_12) -- (w_22) -- (w_32) to[out=0,in=0] (w_12);
    \vertex (r) at (-1,3) {${r}$};
    \draw[] (r) -- (v_1);
    \draw[] (r) -- (v_2);
    \draw[] (r) -- (v_3);
\end{tikzpicture}
}
	\caption{ \label{fig:intGap} An instance of 2NC-TAP such that
	the integrality~ratio of the partition~LP is $\frac{4}{3}$.
	Edges indicated by solid lines have cost~0 and
	edges indicated by dashed lines have cost~1.
	An optimal integer solution picks two of the three unit-cost links
	from each of the cycles $p_1,p_2,p_3,p_1$ and $q_1,q_2,q_3,q_1$.
	An LP solution $\hat{x}$ of cost $3$
	has $\hat{x}_{\ell}=\frac{1}{2}$ for each of the six unit-cost links $\ell$.
}
\end{figure}
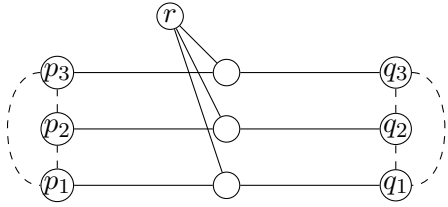
}
{
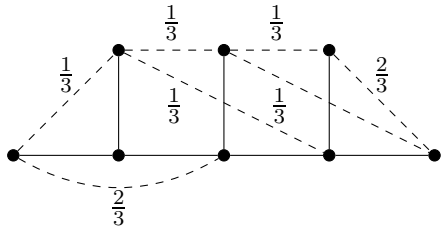
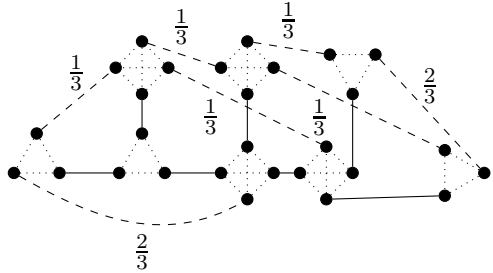
\begin{figure}[htb]
    \begin{subfigure}[t]{0.95\textwidth}
        \centering
        \begin{tikzpicture}[scale=0.70]
            \begin{scope}[every node/.style={circle, fill=black, draw, inner sep=0pt, minimum size = 0.15cm}]
                \node[] (a0) at (0,0) {};
                \node[] (a1) at (2,0) {};
                \node[] (a2) at (4,0) {};
                \node[] (a3) at (6,0) {};
                \node[] (a4) at (8,0) {};
                \node[] (b1) at (2,2) {};
                \node[] (b2) at (4,2) {};
                \node[] (b3) at (6,2) {};
                \draw[] (a0) -- (a1) -- (a2) -- (a3) -- (a4);
                \draw[] (a1) -- (b1);
                \draw[] (a2) -- (b2);
                \draw[] (a3) -- (b3);
            \end{scope}
            \draw[dashed] (a0) -- (b1) node [above, midway] {$\frac{1}{3}$};
            \draw[dashed] (b1) -- (b2) node [above, midway] {$\frac{1}{3}$};
            \draw[dashed] (b2) -- (b3) node [above, midway] {$\frac{1}{3}$};
            \draw[dashed] (b3) -- (a4) node [above, midway] {$\frac{2}{3}$};
            \draw[dashed] (b1) -- (a3) node [below, near start] {$\frac{1}{3}$};
            \draw[dashed] (b2) -- (a4) node [below, near start] {$\frac{1}{3}$};
            \draw[dashed] (a0) to [out=-30,in=-150] (a2);
            \node[] at ($(a1)+(0,-1)$) {$\frac{2}{3}$};
        \end{tikzpicture}
        \caption { \label{f:TAP} An instance of TAP. }
    \end{subfigure}
    \hfill
	\begin{subfigure}[t]{0.95\textwidth}
        \centering
        \begin{tikzpicture}[scale=0.70]
            \node[] (a0) at (0,0) {};
            \node[] (a1) at (2,0) {};
            \node[] (a2) at (4,0) {};
            \node[] (a3) at (5.675,0) {};
            \node[] (a3) at (5.5,0) {};
            \node[] (a4) at (8,0) {};
            \node[] (b1) at (2,2) {};
            \node[] (b2) at (4,2) {};
            \node[] (b3) at (6,2) {};
            \begin{scope}[every node/.style={circle, fill=black, draw, inner sep=0pt, minimum size = 0.15cm}]
	\node[] (a01) at ($(0 : 0.433) + (a0)$) {};
	\node[] (a0b) at ($(90 : 0.75) + (a0)$) {};
	\node[] (a02) at ($(180 : 0.433) + (a0)$) {};
                \draw[dotted] (a01) -- (a0b) -- (a02) -- (a01);
	\node[] (a12) at ($(0 : 0.433) + (a1)$) {};
	\node[] (a1b) at ($(90 : 0.75) + (a1)$) {};
	\node[] (a10) at ($(180 : 0.433) + (a1)$) {};
                \draw[dotted] (a12) -- (a1b) -- (a10) -- (a12);
                \node[] (a23) at ($(0 : 0.5) + (a2)$) {};
                \node[] (a2b) at ($(90 : 0.5) + (a2)$) {};
                \node[] (a21) at ($(180 : 0.5) + (a2)$) {};
                \node[] (a20) at ($(270 : 0.5) + (a2)$) {};
                \draw[dotted] (a23) -- (a2b) -- (a21) -- (a23);
                \draw[dotted] (a2b) -- (a20);
                \draw[dotted] (a21) -- (a20) -- (a23);
                \node[] (a3b) at ($(0 : 0.5) + (a3)$) {};
                \node[] (a31) at ($(90 : 0.5) + (a3)$) {};
                \node[] (a32) at ($(180 : 0.5) + (a3)$) {};
                \node[] (a34) at ($(270 : 0.5) + (a3)$) {};
                \draw[dotted] (a34) -- (a3b) -- (a32) -- (a34);
                \draw[dotted] (a34) -- (a31);
                \draw[dotted] (a3b) -- (a31) -- (a32);
                \node[] (a4b) at ($(0 : 0.5) + (a4)$) {};
                \node[] (a42) at ($(120 : 0.5) + (a4)$) {};
                \node[] (a43) at ($(240 : 0.5) + (a4)$) {};
                \draw[dotted] (a4b) -- (a42) -- (a43) -- (a4b);
                \node[] (b13) at ($(0 : 0.5) + (b1)$) {};
                \node[] (b12) at ($(90 : 0.5) + (b1)$) {};
                \node[] (b10) at ($(180 : 0.5) + (b1)$) {};
                \node[] (b1a) at ($(270 : 0.5) + (b1)$) {};
                \draw[dotted] (b1a) -- (b10) -- (b12) -- (b1a);
                \draw[dotted] (b13) -- (b10);
                \draw[dotted] (b1a) -- (b13) -- (b12);
                \node[] (b24) at ($(0 : 0.5) + (b2)$) {};
                \node[] (b23) at ($(90 : 0.5) + (b2)$) {};
                \node[] (b21) at ($(180 : 0.5) + (b2)$) {};
                \node[] (b2a) at ($(270 : 0.5) + (b2)$) {};
                \draw[dotted] (b2a) -- (b21) -- (b23) -- (b2a);
                \draw[dotted] (b24) -- (b21);
                \draw[dotted] (b2a) -- (b24) -- (b23);
                \node[] (b34) at ($(30 : 0.5) + (b3)$) {};
                \node[] (b32) at ($(150 : 0.5) + (b3)$) {};
                \node[] (b3a) at ($(270 : 0.5) + (b3)$) {};
                \draw[dotted] (b3a) -- (b32) -- (b34) -- (b3a);
            \end{scope}
            \draw[] (a01) -- (a10);
            \draw[] (a12) -- (a21);
            \draw[] (a23) -- (a32);
            \draw[] (a34) -- (a43);
            \draw[] (a1b) -- (b1a);
            \draw[] (a2b) -- (b2a);
            \draw[] (a3b) -- (b3a);
            \draw[dashed] (a0b) -- (b10) node [above, midway] {$\frac{1}{3}$};
            \draw[dashed] (b12) -- (b21) node [above, midway] {$\frac{1}{3}$};
            \draw[dashed] (b23) -- (b32) node [above, midway] {$\frac{1}{3}$};
            \draw[dashed] (b34) -- (a4b) node [above, midway] {$\frac{2}{3}$};
            \draw[dashed] (b13) -- (a31) node [below, near start] {$\frac{1}{3}$};
            \draw[dashed] (b24) -- (a42) node [below, near start] {$\frac{1}{3}$};
            \draw[dashed] (a02) to [out=-30,in=-150] (a20);
            \node[] at ($(a1)+(0,-1.5)$) {$\frac{2}{3}$};
        \end{tikzpicture}
        \caption{\label{f:2NC-TAP} An instance of min-cost 2NCSS.}
    \end{subfigure}
    \caption{ \label{fig:tap}
    		\label{fig:transform}
	\textbf{(a)} An instance of TAP with integrality ratio $\frac{3k+3}{2k+3}$ ($k=3$)
	for the cut~LP relaxation.
	Edges indicated by solid lines have cost~0 and $x$-value~1.
	Edges indicated by dashed lines have cost~1 and
	are labelled with their $x$-values.
\\
	\textbf{(b)} An instance of min-cost 2NCSS
		with integrality ratio $\frac{3k+3}{2k+3}$ ($k=3$)
	for the partition~LP relaxation.
	Edges indicated by solid lines or dotted lines have cost~0 and $x$-value~1.
	Edges indicated by dashed lines have cost~1 and
	are labelled with their $x$-values.}
\end{figure}
}

\clearpage

{

}

\end{document}